\documentclass{PoS}
\usepackage{amsmath}
\usepackage{multicol}
\usepackage{epsfig}
\usepackage{wrapfig}
\usepackage{graphics}
\usepackage{latexsym}
\usepackage{amssymb}
\usepackage{amsmath}
\usepackage{bm}

\renewcommand{\vec}[1]{\boldsymbol{#1}}
\newcommand{\be}{\begin{equation}}
\newcommand{\ee}{\end{equation}}
\newcommand{\ba}{\begin{eqnarray}}
\newcommand{\ea}{\end{eqnarray}}

\newcommand{\Tr}{\,\hbox{\rm Tr}}

\title{Measuring the entropy from shifted boundary conditions}

\ShortTitle{$\dots$ entropy from shifted boundary conditions}

\author{\speaker{Leonardo Giusti}\\
        Dipartimento di Fisica, Universit\`a di Milano-Bicocca,\\
        and INFN, sezione di Milano-Bicocca\\
        Edificio U2, Piazza della Scienza 3\\ 
        20126 Milano, Italy.\\
        E-mail: \email{Leonardo.Giusti@mib.infn.it}}

\author{Michele Pepe\\
        INFN, Sezione di Milano-Bicocca\\ 
        Edificio U2, Piazza della Scienza 3\\ 
        20126 Milano, Italy.\\
        E-mail: \email{Michele.Pepe@mib.infn.it}}

\abstract{We explore a new computational strategy for determining the
equation of state of the SU(3) Yang-Mills theory. By imposing 
shifted boundary conditions, the entropy density 
is computed from the vacuum expectation value of the off-diagonal 
components $T_{0k}$ of the energy-momentum tensor. A step-scaling 
function is introduced to span a wide range in temperature values. 
We present preliminary numerical results for the entropy density and 
its step-scaling function obtained at eight temperature values 
in the range $T_c$--$15~T_c$. At each temperature, discretization 
effects are removed by simulating the theory at several 
lattice spacings and by extrapolating the results to the continuum 
limit. Finite-size effects are always kept below
the statistical errors. The absence of ultraviolet power divergences 
and the remarkably small discretization effects allow for a 
precise determination of the step-scaling function in the explored 
temperature range. These findings establish this strategy as a viable
solution for an accurate determination of the equation of state in a
wide range of temperature values.}

\FullConference{31st International Symposium on Lattice Field Theory 
LATTICE 2013\\
July 29th - August 3rd, 2013\\
Mainz, Germany}

\begin{document}

\section{Introduction}
The standard approach to determine the Equation of State of a gauge theory
on the lattice requires the numerical computation of the free energy 
density~\cite{Boyd:1996bx}. This quantity is usually measured by computing 
its derivative with respect to the bare coupling 
constant via Monte Carlo simulations, and then integrating it back 
analytically. The quartic ultraviolet
divergent term is removed by subtracting the very same quantity at zero 
temperature. Apart from the expansion in the bare coupling constant, this 
strategy requires to accommodate two very different scales 
at the same lattice spacing: the temperature $T$ and the smallest inverse 
correlation length of the zero temperature theory. Only this way discretization
and finite volume effects can be kept under control. Although this approach has 
shown to be successful at low temperature, the zero-temperature simulations 
needed for the subtraction at large temperatures become quickly very expensive. 
This problem prevented numerical computations to access temperatures 
larger than a few $T_c$. The situation improved with the generalization of
the method via the half-temperature subtraction~\cite{Borsanyi:2012ve}. 
However, the expansion in the bare parameters is still needed and the 
simulations remain rather demanding.

In this contribution we present a new computational strategy to avoid the
two-scale problem and, at the same time, the expansion in the bare 
coupling. The entropy density is extracted from the expectation value of a
local operator computed at the desired temperature, and no ultraviolet power
subtractions are needed. 

The basic new ingredient is the use of shifted boundary 
conditions in the temporal 
direction~\cite{Giusti:2010bb,Giusti:2011kt,Giusti:2012yj}. 
In this set up the entropy density can be determined
from the expectation value of the off-diagonal components 
$T_{0k}$ of the energy-momentum tensor~\cite{Giusti:2012yj}, 
a quantity which renormalizes multiplicatively in the SU(3) gauge theory on
the lattice. Finite volume effects are exponentially small
in $(M L)$, where $M$ is the lightest screening mass of the theory and 
$L$ is the linear size in the spatial directions. At large $T$, the 
temperature is the relevant scale in the system: the dominant discretization
effects are proportional to $(a T)^2$, and $M$ is proportional to $T$ 
(more precisely to $(g T)$ or $(g^2 T)$ depending on the temperature). 
Since the observable is local, the computational effort is volume
independent at fixed statistical errors. One can thus simulate
large spatial volumes to account for the small prefactor which enters the
expression of the screening mass, and still keep discretization effects
under control. We test here these ideas in the SU(3) Yang--Mills theory.
They can, however, be easily generalized to gauge theories with fermions like
QCD.

\section{Entropy density from shifted boundary conditions}
The SU(3) Yang--Mills theory  at finite temperature can be formulated in the 
Euclidean path integral formalism by imposing on the fields 
periodic boundary conditions in the compact direction up to 
a shift $\vec\xi$ in the spatial 
direction~\cite{Giusti:2010bb,Giusti:2011kt,Giusti:2012yj}
\begin{equation}
A_\mu (L_0,\vec x) = A_\mu (0,\vec x - L_0\, \vec \xi),
\end{equation}
where $A_\mu$ is the gauge field and $L_0$ is the time extension. 
In the thermodynamic limit, the invariance 
of the dynamics under the SO(4) group implies that the free energy 
density $f(L_0 ,\vec \xi)$  satisfies
\begin{equation}
f(L_0 ,\vec \xi) = f(L_0 \sqrt{1+\vec \xi ^2},{\vec 0}).
\end{equation}
Thus the free energy depends on the length of the compact 
direction $\beta= L_0 \sqrt{1+\vec \xi^2}= T^{-1}$
which fixes the inverse temperature of the system, while
it is independent on its orientation with respect to 
the space directions. This redundancy 
implies that the total energy and momentum distributions of the 
thermal theory are related, and interesting Ward identities (WIs) 
follow. In particular, one obtains that 
\begin{equation}
\langle T_{0k} \rangle _{\vec \xi} = \frac{\xi_k}{1- \xi_k ^2} 
\left[ \langle T_{00} \rangle _{\vec \xi} - \langle T_{kk} \rangle _{\vec \xi} \right]\; , 
\end{equation}
where $\langle \cdot \rangle _{\vec \xi}$ stands for the vacuum expectation 
value 
computed in the thermal quantum system with shift ${\vec \xi}$, and  
$T_{\mu\nu}$ is the energy-momentum tensor of the theory. For a 
non-vanishing shift, 
it then follows that the entropy density $s(T)$ is given by
\begin{equation}\label{eq:sxi}
\frac{s(T)}{T^3} =  -\frac{L_0^4 (1+\vec \xi ^2)^3}{\xi_k} 
\langle T_{0k} \rangle _{\vec \xi}\; ,\qquad T=\frac{1}{L_0 \sqrt{1+\vec \xi^2}}\; .   
\end{equation}
Remarkably, the entropy density can be obtained directly from the vacuum 
expectation value of the off-diagonal component $T_{0k}$ of the 
energy-momentum tensor which does not vanish since the shift softly 
breaks the parity symmetry. Based on the Eq.~(\ref{eq:sxi}), 
a step-scaling function $\Sigma (T,r)$ for the normalized entropy 
density can be defined as 
\begin{equation}\label{stepfun_cont}
\Sigma (T,r) = \frac{s(T')/T'^3}{s(T)/T^3} = 
\frac{(1+{\vec \xi'} ^2)^3\,\, \xi_k}{(1+\vec \xi ^2)^3\,\, \xi'_k} 
\frac{\langle T_{0k} \rangle _{\vec \xi'} }{\langle T_{0k} \rangle _{\vec \xi} }\; , 
\end{equation}
where $\vec \xi$ and $\vec \xi'$ are two different shifts. Since $L_0$ is held 
fixed, the step $r$ in the temperature is given by the ratio 
$r=T'/T= \sqrt{1+\vec \xi ^2}/\sqrt{1+{\vec \xi'} ^2}$. Following the 
approach in Ref.~\cite{Luscher:1993gh}, the entropy density at a given 
temperature can then be obtained by solving the recursion relation
\be
v_0=\frac{s(T_0)}{T^3_0}\; , 
\qquad v_{k+1} = \Sigma_s(T_k,r)\, v_k\; , \qquad T_k=T_0\, r^k\; ,
\ee
once the entropy density $v_0$ is computed at temperature $T_0$.

\section{Entropy density on the lattice}
We set up the SU(3) Yang--Mills theory on a four-dimensional 
lattice of size $ L_0 \times L^3$ and spacing $a$ by discretizing
the gluons with the standard Wilson plaquette action. We impose 
periodic boundary conditions in the spatial directions and 
shifted boundary conditions along the compact direction
\begin{equation}
U_\mu(L_0,\vec x) = U_\mu(0,\vec x- L_0\, \vec \xi)\; ,
\end{equation}
where $\vec \xi $ is the shift vector and $U_\mu(x)$ 
are the gauge links. We consider the clover 
formulation of the energy-momentum tensor on the 
lattice~\cite{Caracciolo:1989pt}
\be\label{eq:TmunuLat}
T_{\mu\nu} =  \frac{\beta}{6}\Big\{F^a_{\mu\alpha}F^a_{\nu\alpha}
- \frac{1}{4} \delta_{\mu\nu} F^a_{\alpha\beta}F^a_{\alpha\beta} \Big\}\; ,
\ee
where $\beta=6/g_0^2$, and $g_0$ is 
the bare coupling constant. The field strength tensor is 
defined as 
\be
F^a_{\mu\nu}(x) = - \frac{i}{4 a^2} 
\Tr\Big\{\Big[Q_{\mu\nu}(x) - Q_{\nu\mu}(x)\Big]T^a\Big\}\; ,  
\ee
where
\ba
Q_{\mu\nu}(x) & = & 
U_\mu(x)\, U_\nu(x+ a\hat\mu)\, U^\dagger_\mu(x + a\hat\nu)\,U^\dagger_\nu(x)\nonumber\\[0.125cm]
& + & U_\nu(x)\, U^\dagger_\mu(x-a\hat\mu+a\hat\nu)\, U^\dagger_\nu(x - a\hat\mu)\,
U_\mu(x-a\hat\mu)\\[0.125cm]
& + & U^\dagger_\mu(x-a\hat\mu)\, U^\dagger_\nu(x-a\hat\mu-a\hat\nu)\, 
U_\mu(x-a\hat\mu-a\hat\nu)\,U_\nu(x-a\hat\nu)\nonumber\\[0.125cm]
& + & 
U^\dagger_\nu(x-a\hat\nu)\, U_\mu(x-a\hat\nu)\, 
U_\nu(x + a\hat\mu - a\hat\nu)\, U^\dagger_\mu(x)\; . \nonumber
\ea
\begin{wraptable}{r}{0.45\textwidth}
\vspace{-0.675cm}

\small
\begin{tabular}{ccccc}
\hline
$T/T_0$  &$L/a$&$L_0/a$&$\beta$&$TL$\\
\hline
$1/\sqrt{2}$  &$80$ &$3$& 6.0403 &$13.4$\\     
$1/\sqrt{2}$  &$96$ &$3$& 6.0403 &$22.6$\\     
$1/\sqrt{2}$  &$96$ &$4$& 6.2257 &$12.0$\\
$1/\sqrt{2}$  &$128$&$5$& 6.3875 &$12.8$\\
\hline
$1$  &$80$ &$3$& 6.2670 &$13.4$\\     
$1$  &$96$ &$3$& 6.2670 &$22.6$\\     
$1$  &$96$ &$4$& 6.4822 &$12.0$\\
$1$  &$128$&$5$& 6.6575 &$12.8$\\
\hline
$\sqrt{2}$  &$80$ &$3$& 6.5282 &$13.4$\\     
$\sqrt{2}$  &$96$ &$3$& 6.5282 &$22.6$\\     
$\sqrt{2}$  &$96$ &$4$& 6.7533 &$12.0$\\
$\sqrt{2}$  &$128$&$5$& 6.9183 &$12.8$\\
\hline
$2$  &$80$ &$3$& 6.7791 &$13.4$\\     
$2$  &$96$ &$3$& 6.7791 &$22.6$\\     
$2$  &$96$ &$4$& 7.0201 &$12.0$\\
$2$  &$128$&$5$& 7.2068 &$12.8$\\
\hline
$2\sqrt{2}$  &$80$ &$3$& 7.0694 &$13.4$\\     
$2\sqrt{2}$  &$96$ &$3$& 7.0694 &$22.6$\\     
$2\sqrt{2}$  &$96$ &$4$& 7.3100 &$12.0$\\
$2\sqrt{2}$  &$128$&$5$& 7.4963 &$12.8$\\
\hline
$4$  &$80$ &$3$& 7.4120 &$13.4$\\     
$4$  &$96$ &$3$& 7.4120 &$22.6$\\     
$4$  &$96$ &$4$& 7.6541 &$12.0$\\
$4$  &$128$&$5$& 7.8435 &$12.8$\\
\hline
$4\sqrt{2}$  &$80$ &$3$& 7.7039 &$13.4$\\     
$4\sqrt{2}$  &$96$ &$3$& 7.7039 &$22.6$\\     
$4\sqrt{2}$  &$96$ &$4$& 7.9489 &$12.0$\\
$4\sqrt{2}$  &$128$&$5$& 8.1405 &$12.8$\\
\end{tabular}
\caption{The parameters used in the numerical study. For each parameter set, 
simulations with two shifts $\vec \xi' = (1,0,0)$ and 
$\vec \xi = (1,1,1)$ have been carried out. This corresponds to the two 
temperatures:  $T = (2\, L_0)^{-1}$ and $T' = (\sqrt{2}\, L_0)^{-1}$.
\label{tablepar}}
\vspace{-0.75cm}

\end{wraptable}
On the lattice translational invariance is broken down to a discrete 
sub-group, and the energy-momentum tensor has to be renormalized.
The momentum density renormalizes multiplicatively 
$T_{0k}^{\rm R} = Z_{_T} T_{0k}$, and its renormalization constant $Z_T$
is fixed by imposing suitable WIs \cite{Giusti:2011kt,Giusti:2012yj}. 
As a consequence $Z_T$ depends only on the bare coupling constant and,
up to discretization effects, it is independent on the volume, 
the temperature, the shift parameter, etc. Ultimately which WIs 
and/or kinematics yield the most accurate result must be investigated 
numerically. The factor $Z_T$ cancels out in the 
lattice definition of the step-scaling function
\begin{equation}\label{stepfun_latt}
\Sigma (T,r) = 
\frac{(1+{\vec \xi'}^2)^3\,\, \xi_k}{(1+{\vec \xi} ^2)^3\,\, \xi'_k} 
\frac{\langle T_{0k} \rangle _{\vec \xi'} }{\langle T_{0k} \rangle _{\vec \xi} }  
\end{equation}
which has a universal continuum limit as it stands. In the remaining part
of these proceedings we will focus on the numerical determination of the 
step-scaling function.

\section{Numerical~computation}
\begin{figure}[thb]
\begin{center}
\vspace{0.5cm}

\includegraphics[width=0.65\textwidth]{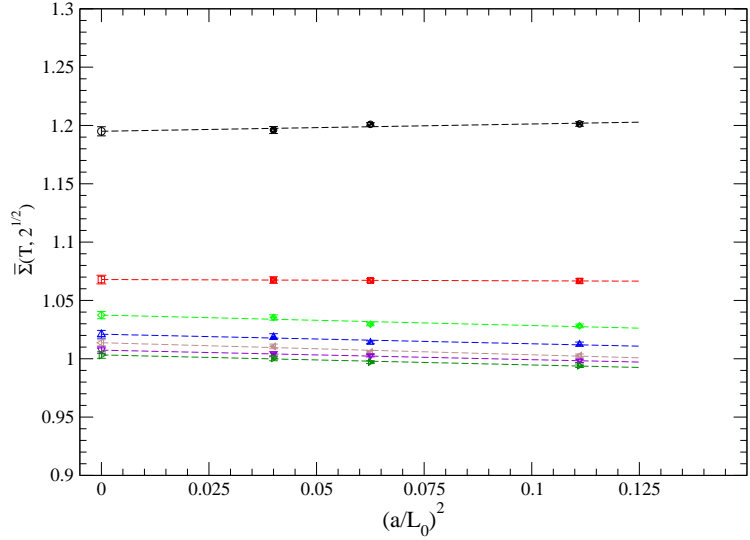}
\caption{Continuum limit extrapolation of the entropy step-scaling function 
${\overline \Sigma} (T,\sqrt{2})$ at the 7 temperatures listed in 
Table~\protect\ref{tablepar}. The temperature increases from top to bottom. 
The dashed lines correspond to linear fits of the numerical data.
\label{fig:CL}} 
\end{center}
\end{figure}
We have simulated the SU(3) Yang--Mills theory by sweeping the lattice
with 1 heatbath and 3 over-relaxation updates of all link variables. 
At each value of $\beta$ and $L_0/a$, we have measured 
$\langle T_{0k} \rangle _{\vec \xi}$ for two values of the shift, 
$\vec\xi' = (1,0,0)$ and $\vec\xi = (1,1,1)$, corresponding to a step 
of $r=\sqrt 2$ in the temperature. In order to extrapolate 
$\Sigma (T,\sqrt{2})$ to the 
continuum limit, at each temperature we have collected data at three 
different values of the lattice spacing $L_0/a=3$, $4$, and $5$; runs at 
$L_0/a=6$ are currently in progress to have a better control on the 
systematics due to the extrapolation. We have measured 
$\Sigma (T,\sqrt{2})$ 
at 7 temperatures in the range $T_0/\sqrt{2}\,$--$\,4\sqrt{2}\, T_0$, with 
the values separated by steps of about $\sqrt{2}$ to match the value of $r$. 
Our reference temperature  has been fixed to $T_0=L^{-1}_{\rm max}$, where 
$L_{\rm max}$ is taken from Ref.~\cite{Capitani:1998mq}. It corresponds to 
$T_0\simeq 1.802 T_c$, where $T_c$ is the critical temperature 
computed in Ref.~\cite{Lucini:2003zr}. In the first three steps 
($k=-1,0,1$) the value of $\beta$ of each run has been fixed from
$r_0/a$, by requiring that $L_{\rm max}/r_0=0.738(16)$~\cite{Necco:2001xg}. 
For each pair of steps 
$k=2j,\, 2j+1$, with $j\geq 1$, we interpolate quadratically in 
$\ln{(L/a)}$ each set of data at 
constant $\bar g^2(L_j)$ in Table A.1 of Ref.~\cite{Luscher:1993gh} 
supplemented by the corresponding data in Table A.4
\be
\beta = a_j + b_j \ln{\Big(\frac{L}{a}\Big)} + c_j 
\ln^2{\Big(\frac{L}{a}\Big)}\; ,
\ee
and we fix $\beta$ by requiring that $T_{2j}\, a = a/L_j$
and $T_{2j+1}\, a =\sqrt{2}\, a/L_j$. The $\beta$ 
values obtained are reported in Table~\ref{tablepar},
together with the spatial and time extent of the lattices simulated.
To keep finite volume effects below the statistical errors, we have chosen 
$T L \geq 12$. At each coarser lattice spacing, we have also simulated a 
smaller 
volume to verify that finite volume effects are below the statistical errors. 
Having large lattices does not increase the cost, since the latter is volume
independent thanks to the locality of the observable.

\section{Results and conclusions}
\begin{figure}[thb]
\begin{center}
\vspace{0.5cm}

\includegraphics[width=0.65\textwidth]{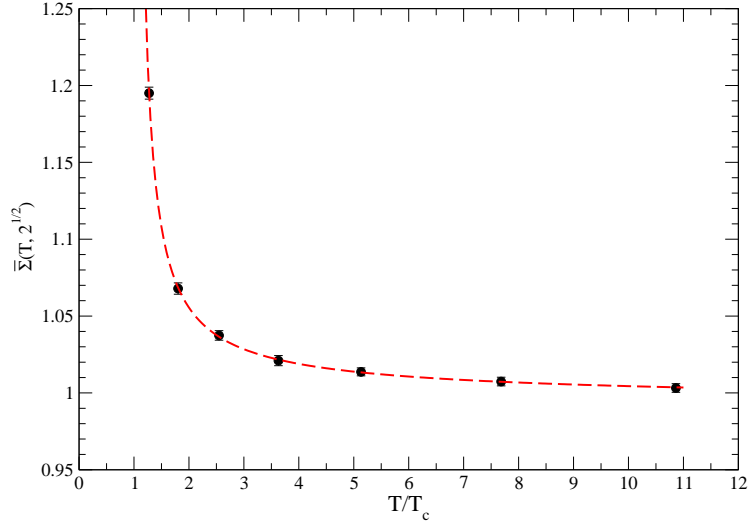}
\caption{The temperature dependence of the entropy step-scaling function in the 
continuum limit. Error bars take into account statistical errors only.
The dashed line is an interpolation of the points to guide 
the eyes.
\label{fig:step}}
\end{center}
\end{figure}

\begin{figure}[thb]
\begin{center}
\vspace{1.0cm}

\includegraphics[width=0.65\textwidth]{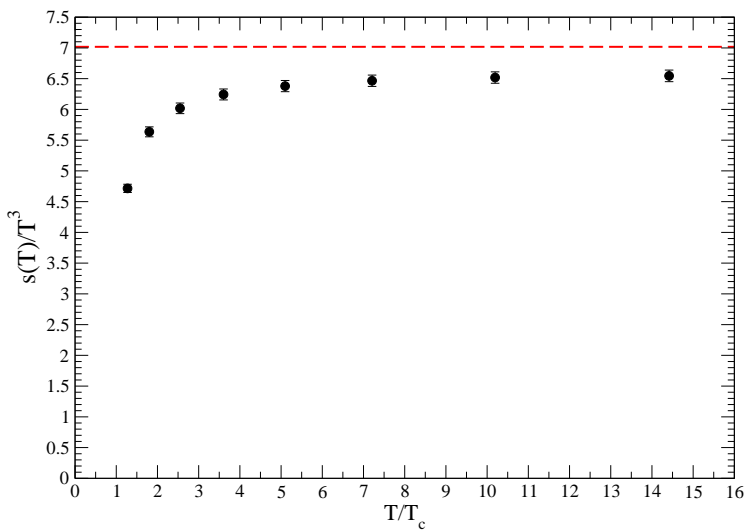}
\caption{The normalized entropy density as a function of $T/T_c$. The dashed 
line is the Stefan-Boltzmann result $s_{\rm SB}/T^3=32\pi^2/45$.\label{entropy}}
\end{center}
\end{figure}

The step-scaling function in the free theory, $\Sigma_0(T,\sqrt{2})$, can 
be computed analytically on the lattice. It has 
small discretization effects of the order of a few \% in the range 
of $L_0/a$ we are interested in~\cite{Giusti:2012yj}. 
This fact is confirmed in the interacting theory by our preliminary results 
shown in Fig.~\ref{fig:CL}, where the difference  
\be
{\overline \Sigma}(T,\sqrt{2})=\Sigma(T,\sqrt{2}) - (\Sigma_0(L_0)-1)\; , 
\ee
is plotted as a function of $(a/L_0)^2$. The residual discretization effects 
in ${\overline \Sigma}(T,\sqrt{2})$ are at the per-mille level already at 
$L_0/a=3$.  The linear extrapolation in $(a/L_0)^2$ provides a satisfactory 
fit of the numerical data, as shown by the dashed lines in the plot. 
When the data at $L_0/a=6$ will become available, we 
will be able to quantify also the systematics due to the continuum 
limit extrapolation. At that point it may be useful to attempt a global 
fit of all data points by parameterizing the coefficient of the 
discretization effects. In Fig.~\ref{fig:step} we plot the continuum 
limit values of the step-scaling function versus the temperature, where
the errors are statistical only. The temperatures at which the
step-scaling function has been measured are 
not always precisely related by a factor $\sqrt{2}$. However, by interpolating 
the results of the step-scaling function, the step $\sqrt{2}$ in the 
temperature can be accurately enforced and the temperature dependence 
of $s(T)/T^3$ can be reconstructed. For the sake of the presentation, in the 
Fig.~\ref{entropy} we show the temperature dependence of $s(T)/T^3$ obtained
from the entropy step-scaling function by fixing the overall normalization from
the results for the entropy density at $T=4.1 T_c$ published in 
\cite{Giusti:2010bb}. The statistical uncertainty on that measurement 
dominates the error bars. We expect to improve significantly the accuracy of 
these results once $Z_T$ will be fixed from the WIs. A first step in this 
direction has already been taken at this conference~\cite{Robaina:2013zmb}. 

Our preliminary results establish the strategy followed in this work 
as a viable and efficient solution for determining accurately the equation
of state in a wide range of temperature values. The cost of these simulations
is moderate, as proved by the fact that all computations presented here 
have been done using a few millions of core hours on a BG/Q. Simulations 
to properly quantify the systematic errors due to the continuum limit 
extrapolation are underway. 

We thank H. B. Meyer and D. Robaina for discussions. Simulations have been 
performed on Blue Gene/Q at CINECA (CINECA-INFN agreement) and on the Turing
machine at the University of Milano-Bicocca. We thank these Institutions for
the generous allocation, the support and the technical help. This work was
partially supported by the MIUR-PRIN contract 20093BMNNPR and by the INFN
SUMA project.

\end{document}